# A System Architecture for Software-Defined Industrial Internet of Things

Peng Hu
CMC Microsystems
Kingston, Ontario, Canada K7L 3N6
Hu@cmc.ca

*Abstract*—Wireless sensor networks have been a driving force of the Industrial Internet of Things (IIoT) advancement in the process control and manufacturing industry. The emergence of IIoT opens great potential for the ubiquitous field device connectivity and manageability with an integrated and standardized architecture from low-level device operations to high-level data-centric application interactions. This technological development requires software definability in the key architectural elements of IIoT, including wireless field devices, IIoT gateways, network infrastructure, and IIoT sensor cloud services. In this paper, a novel software-defined IIoT (SD-IIoT) is proposed in order to solve essential challenges in a holistic IIoT system, such as reliability, security, timeliness scalability, and quality of service (QoS). A new IIoT system architecture is proposed based on the latest networking technologies such as WirelessHART, WebSocket, IETF constrained application protocol (CoAP) and software-defined networking (SDN). A new scheme based on CoAP and SDN is proposed to solve the QoS issues. Computer experiments in a case study are implemented to show the effectiveness of the proposed system architecture.

## I. INTRODUCTION

Over the recent decade, the technological advancement in wireless sensor networks (WSNs) and embedded systems has enabled numerous applications in different domains including Internet of Things (IoT) systems [1]. A special class of IoT-enabled industrial production systems is called the Industrial Internet of Things (IIoT), which provides efficacy and economic benefits to system installation, maintainability, scalability, and interoperability. The holistic architecture of IIoT systems and the broad spectrum of IIoT applications require a flexible software definable capability in order to solve the essential challenges, including timeliness, security, reliability, scalability, and quality of service (QoS), where few of them have been addressed by the existing IoT technologies which hardly provide a solution to the challenges in the key IIoT architectural elements, such as wireless field devices (FDs), gateways (GWs), and sensor cloud (SC) services with the controllability over the network infrastructure devices as a holistic system.

FDs are important wireless sensor networking devices which acquire and react to field information in the industrial applications. These devices can be connected by open wireless standards but a challenge is how to efficiently interoperate with the low-power FDs which need to be used in the industrial environment for many years of operation. The WirelessHART (i.e. IEC 62591) and ISA100.11a standards [2], [3] aiming to address this challenge are popular in manufacturing automation systems. However, systems powered by these standards are difficult to accommodate various heterogeneous FDs and are difficult to extend and to maintain. Therefore, an interoperable application-layer communication protocol for different IIoT devices should be made available. In order to provide flexible interoperation to scale-up IIoT systems and to accommodate various underlying field network technologies, we need a software-defined management solution based on an open interoperation protocol.

A GW is an important architectural element which needs to provide seamless connectivity between one or more FDs and a GW (i.e., FD-GW) and between a GW and an SC (i.e., GW-SC) with proper remote configurability in the middleware tier. There are several possible solutions to the IIoT middleware framework. The traditional Common Object Request Broker Architecture (CORBA) system is a viable option but it is complex to implement, compared to the HTTP-based web service technology preferred in many distributed systems [4], because the web service based solutions can be decoupled from the central services and access the machine data with the GW middleware. The OPC unified architecture (OPC UA) specifications define the transmission control protocol (TCP)-based connectivity from traditional FDs to the programmable logic controller (PLC) and distributed control systems (DCS) as well as high-level web services. However, this solution mainly targets the process control devices and not the low-power wireless sensor nodes, and it lacks the flow optimization for the underlying network-layer packets and QoS guarantees, which can be addressed by software-defined networking (SDN) technologies. The real-time data transmission capability of GW-SC can be brought by the new HTML5 WebSocket or CoAP [5] protocols where the equally important connectivity of FD-GW can be brought by the CoAP-based protocol. In this way, the software-defined capacity can be added to each of the IIoT architectural elements.

In this paper, a new software-defined IIoT (SD-IIoT) architecture with its key components based on the low-power industrial WSNs will be proposed and the software-defined functions for FDs, GWs, and SC based on WebSocket, CoAP, and SDN technologies will be discussed. The paper is organized as follows: the related work is discussed in Section II; Section III discusses the new software-defined IIoT architecture in the scale-up IIoT systems; Section IV shows the effectiveness of the proposed system architecture with experimental results; and the conclusive remarks and future work are presented in Section V.

## II. RELATED WORK

The state-of-the-art works in the literature and industry for interoperable IIoT systems have been driving the development of the three essential architectural elements in the IIoT systems: FDs, GWs, and SCs. FDs with sensors and/or actuators are important field networking devices functioning as the intrinsic data source of the IIoT systems, while GWs and SCs play an important role in the data connectivity and flexibility and they are responsible of system scalability and transferring the little data to big data which can then be processed with data analytics for business logics.

There are several mainstream low-power FD technologies that can be used in an IIoT system, such as ZigBee, 6LoWPAN, IETF 6tisch, WirelessHART, ISA100.11a, as well as classical WSAN [6] and M2M systems. Generally, they can be used in a field WSN as part of an IIoT system, but each of them focuses on different markets and application domains. For example, ZigBee and 6LoWPAN mainly target at wireless control and condition monitoring application such as building/home automation and smart energy systems; WirelessHART and ISA100.11a target at process control and manufacturing industry with the time division multiple access (TDMA)-based communication and channel-hopping scheme for robustness and data bandwidth [3]; and IETF 6tisch is a work in progress which realizes the IEEE 802.15.4e time slotted channel hopping (TSCH), where an improved IEEE 802.15.4e media access control (MAC) sublayer for low latency deterministic networks (LLDNs) is introduced in [7]. In fact, there is no one-size-fits-all solution and the technology fragmentation also limits the scalability of IIoT systems.

There are different SC services providing essential connectivity to the geographically distributed sensing devices. In recent years, the number of IoT cloud service providers has been increasing. Small FDs have to connect to a central service in order to create various application for end users. SC services generally provide data storage, data visualization, data connectivity, and data analytics. For examples, SensorCloud and ThingWorx provide cloud services to the users with simplified data operation APIs, where users can easily upload data to the services with a RESTful interface. Amazon Kinesis can process the at-scale streaming data cloud service, and IBM Bluemix provides similar cloud and data analytics services. However, these data services are mainly for general-purposes consumer applications while they hardly address the real-time capability, such as data synchronization and real-time data visualization between IoT devices and remote services. Moreover, these SCs hardly support the automation and manufacturing applications directly, which makes them difficult to be applied in many IIoT scenarios.

With remote SC services, the fast data analytical tasks can be completed by high-end servers or a high-performance computing cluster (HPCC), so the question is how to ensure the GW-SC data synchronization for real-time data visualization, processing, and storage. The existing solutions use a local server between IoT devices and remote server to store a big chunk of FD data and then upload them to a remote server, which can result in a couple of severe problems: 1) the time of device data stored on the server is not synchronized with the device data generation time; and 2) the real-time data analysis is impossible on the server end. This solution hinders the IIoT from being a real-time data-centric system, in particular for safety-critical IIoT systems. Therefore, we need to have an efficient solution that can minimize and adapt timeliness and responsiveness requirements in the IIoT-class data-centric environment.

In addition to the SC services, another important element is the GW with middleware employing efficient message exchange protocols that can process and exchange the information with FDs. It provides essential interoperability for FD-GW and GW-SC connections and extensibility to heterogeneous systems. HTML5 WebSocket, CoAP, and MQTT-S are the mostly popular message exchange protocols. WebSocket uses HTTP over TCP which can provide bidirectional connectivity between IoT middleware modules and FDs; MQTT-S provides two-way communication capability over the user datagram protocol (UDP) for sleepy sensor networks and it is messaging oriented; and CoAP is an IETF standard which has web service oriented architecture and provides additional services than MQTT-S, and it has open-source and commercial realizations, such as ARM Sensinode/mbed and Californium CoAP. We will extend the CoAP in order to make it work in the proposed FD-GW connections and WebSocket in the GW-SC connections.

## III. SOFTWARE-DEFINED IIOT SYSTEMS

### A. Controllability over IIoT Systems

The controllability over the IIoT system relies in three factors: (a) the control over a field WSN, (b) the control over the internal and external backbone network infrastructure, (c) the control over the software applications and networking protocols. Having sufficient control over the three factors will improve the timeliness, reliability, and QoS.

The control over a field WSN is related to a chosen wireless technology and it is comparably easy to guarantee the timeliness over the wireless channels. For example, the TDMA-based scheduling scheme used in WirelessHART protocol can have deterministic latency of the packet transmission from FDs to a gateway or an access point. The control over the local and external networks is related to the network medium and infrastructure devices, which are hardly to be controlled easily. The delay caused by software applications is application-specific and it should be optimized with the first two factors mentioned above. There are usually several different kinds of network delays, such as processing delay, queuing delay, transmission delay, and propagation delay. The processing and queuing delays relate to the routers and switches in the network, which are controllable and optimizable using SDN technologies. The transmission delay means the time to put the packets onto the link, which can be considered as a constant with a short duration. The propagation delay is the time a signal takes to a destination which relates to the transmission medium and is usually a constant. Therefore, in this paper, we will focus on the discussion on the network controllability.

### B. Software-Defined IIoT (SD-IIoT) System

A solution to dynamic controllability is providing the IIoT system with the software definable capability. An efficient data-centric IIoT system needs the real-time performance which can be achieved by adding dynamic controllability to its key architectural elements. A software-defined approach is proposed where an IIoT system has the control over the configuration parameters and data in the architectural elements. There are disparities between the proposed SD-IIoT and the SDN or software-defined WSN [8], as the SD-IIoT emphasizes the holistic application-specific performance of FD-GW and GW-SC connections and interoperable. In this case, CoAP-based networking protocols are used to address heterogeneity of underlying FD and GW technologies and the RESTful interface and WebSocket are used to address the GW-SC communications and data transmissions. Specifically, an SD-IIoT system consists of the following elements:

- IIoT FD: it is the wireless sensing device compatible with specific industrial standardized networking protocols and can be controlled with an application control interface.
- IIoT GW: it provides configurability in bidirectional FD-GW connections. The control plane components can be resided in the GW or the host connected to the GW. The data manageability of the GW includes the data processing schemes, data adapters, protocol converters.
- IIoT SC: it provides configurability for the high-level IIoT application hosted by SC services. The real-time QoS performance between GW and SC is guaranteed by the SC control plane.

*C. System Architecture*

The proposed architecture of an SD-IIoT system is depicted in Fig. 1, where the IIoT SC, IIoT GW, and IIoT FD are connected with the main components in the vertical control plane and data plane. The term IIoT GW and GW are used interchangeably in the rest of the paper.

For the IIoT SC, there are two main modules: controller and data manager. The controller module in the control plane is responsible of QoS configuration for GW-SC data flows, FD configuration for FD and field WSN parameters (i.e., network formation, device duty cycle, radio output power, etc.), SDN control through the OpenFlow SDN controller with networking resources, and security configuration for GW-SC and FD-GW data transmissions with the security controller. The QoS controller in the IIoT SC defines and stores QoS policies for the network backbone and the field WSN which is realized by the QoS controller in the IIoT GW. The data manager module provides basic data management services, such as data processing/analytics, data storage, data visualization, and device management. The GW-SC and FD-GW network maintenance data are managed by the maintenance data management module. For example, GW-SC maintenance data can be obtained by using the simple network management protocol (SNMP) protocol, while the FD-GW maintenance data can be obtained by the WSN diagnostic messages. The application interfaces provide end users with the essential IIoT configuration and application services. For example, an SDN QoS policy is defined through the RESTful interfaces defined by JavaScript Object Notation (JSON) files.

The IIoT GW in Fig. 1 has a controller module implementing control plane functions and a data manager module implementing data plane functions for data processing, adaptation and conversion, where the controller module contains a QoS controller which is controlling the field WSN and resources defined by the QoS controller component in the IIoT SC. The QoS controller can also autonomously determine the data flow scheduling corresponding to dynamic network conditions, and use the SDN controller interface to find the best route for the real-time data transmissions. Other field network configurations such as topology and data update intervals are controlled by the field network controller. Data synchronization controller component is responsible of making sure the timeliness requirements of GW-SC data transmissions. Because the traditional HTTP connection between clients and servers are uni-directional and much transmission delay are caused by the chunky HTTP packet sizes. Instead, we use WebSocket bi-directional connection to directly represent the data, where the time-stamped data can be opted to be stored in the database. Security schemes of FD-GW and GW-SC data transmissions can be configured by the security controller component. For example, datagram transport layer security (DTLS) can be configured to be used for FD-GW communication with CoAP, and transport layer security (TLS) can be used for GW-SC communication with WebSocket. The FD-GW data adapter module is responsible of data conversion between FDs and GW. For example, if WirelessHART is used on the FDs, CoAP-HART IP adapter is required to translate the CoAP command packets into HART-IP commands which can finally control the WirelessHART-based WSN. The protocol converter is responsible of conversion between different industrial communication protocols such as HART IP and OPC, where an HART IP-OPC protocol adapter can provide interoperability between HART systems and OPC systems. The GW-SC data adapter is doing the data conversion between the SC and GW. The WebSocket-CoAP adapter is an example of this GW-SC adapter which can perform protocol translation functions between the WebSocket and CoAP packets. The FD device adapter works as an abstract interface for heterogeneous underlying FD networks, such as WirelessHART, ISA100.a, 6tisch, and ZigBee. The FD device adapter for different networks provides essential scalability for an IIoT system.

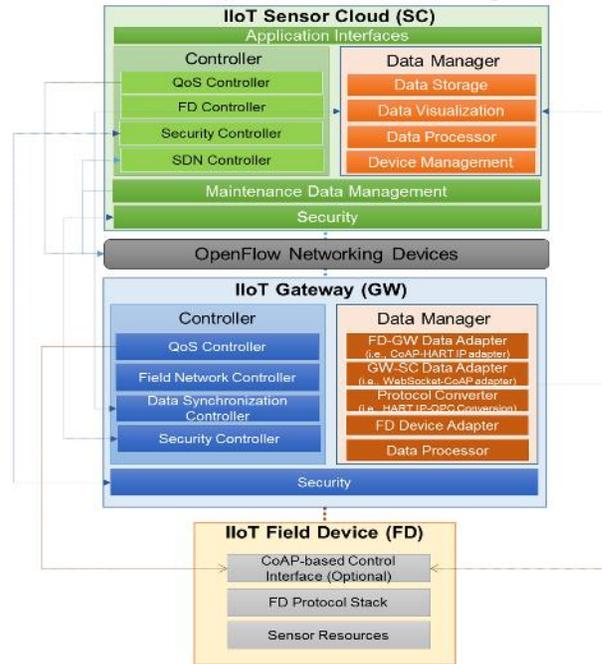

Fig. 1. SD-IIoT system architecture with essential components in the architectural elements. The interaction relationships between components are shown in dotted lines.

*D. QoS Configuration with SDN Controllers*

The SDN controller is used to control the data plane of the multilayer open virtual switches (OVS) in the controlled industrial network. Each border router representing an IIoT gateway is connected to four distributed OVSes as shown in Fig. 2. A Floodlight controller is running on top of the OVS which is controlled by the IIoT SC server. The field WSN consists of FDs connected to the GW or border router as the GW-SC communication is through the IP networks.

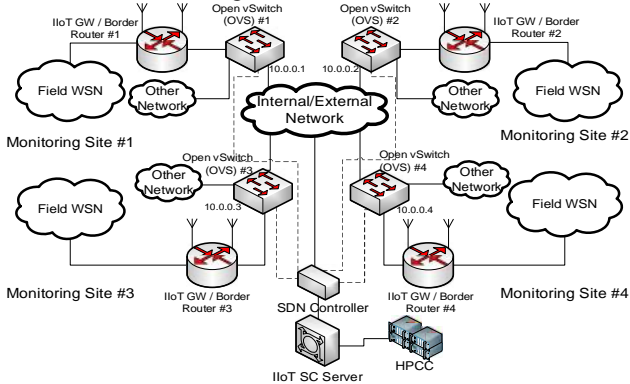

Fig. 2. Network deployment diagram of the scalable SD-IIoT system

We propose to use the CoAP as a communication protocol for exchanging QoS policies among the key architectural FD-GW QoS components such as GW data processor, GW QoS controller, and FDs, where a CoAP server endpoint is deployed on the FDs while a CoAP server endpoint and a WebSocket server endpoint are deployed on the IIoT GW data processor. The WebSocket client endpoint on the IIoT SC QoS controller can therefore interoperate with IIoT GW and FDs.

In TABLE I, different types of industrial applications with different timing requirements are shown, where depending on generic industrial sensing applications, we can further categorize the sensor classes into three QoS groups. Then we need to see for each QoS group what the traffic pattern is and how the tolerance for packet loss, delay, and jitter looks like. For the QoS group 1, the sensing data have small fixed-sized packets, constant emission rate, and inelastic low-rate flows, as well as very low tolerant to packet loss, delay, and jitter. For QoS group 2, the sensing data may allow some variable sizes and inelastic and variable data rate, as well as very low tolerance for packet loss and delay. For the QoS group 3, the traffic is similar to the QoS group 3, and still has very low tolerance for packet loss but low tolerance for delay and jitter. From [9], we should use differentiated services codepoint (DSCP) Express Forwarding (EF) for QoS group 1, CS4 for QoS group 2 with real-time data flow transmissions, CS4 or AF21-23 for QoS group 3 depending on the monitoring applications. In addition, for the network control flows that transmit the policy and QoS controller messages, we allow the jitter but low packet delay and loss with variable sizes and they use DSCP CS6. In addition, the QoS policy should also check the flow priorities dependent on the dynamic traffic controlled by the network control commands (e.g., commands for requiring more frequent updates on sensing data reporting). A database in an HPCC in Fig. 2 can be provided as a repository of the network management data with SNMP in order to monitor the CoAP and HTTP packet transmissions. In Fig. 2, each monitoring site has one GW and one OVS that can be remotely configured by the SDN controller on the IIoT SC QoS controller.

TABLE I. TYPICAL TIMING REQUIREMENTS OF APPLICATIONS [2] AND QOS IN THE PROCESS AUTOMATION DOMAIN

| ISA-100.11a Sensor Classes | Applications | Update Freq. | QoS Group | DSCP |
|---|---|---|---|---|
| Class 5 Class 4 | Monitoring & supervision | secs-days | Group 3 | CS4 or AF21-23 |
| | Temperature sensor | 5 s | | |
| | Pressure sensor | 1 s | | |
| Class 3 Class 2 | Closed loop control | 10-500 ms | Group 2 | CS4 |
| | Pressure sensor | 10-500 ms | | |
| | Temperature sensor | 500 ms | | |
| Class 1 Class 0 | Interlocking and Control | 10-250 ms | Group 1 | EF |
| | Motor | 10-250 ms | | |
| | Valve | 10-250 ms | | |

*E. Expected Time Span of WebSocket/CoAP Messages*

The expected time span $t_e$ of a confirmable WebSocket/CoAP message is: $T * ((2^C) - 1) * F$, where $T$ is the timeout of the ACK message, and $C$ is the maximum retransmit times of a confirmable message, and $F$ is a random factor. If the values of $T$, $C$, and $F$ are 2 ms, 4, and 1.5, the value of $t_e$ is 45 ms. In this case, $t_e$ provides an indicator of the worst-case scenario. Later, we will see that in the controlled environment the actual $t_e$ is quite small as $T$ and $C$ can be kept quite small.

IV. EXPERIMENTAL RESULTS

In this section, we will discuss about the important IIoT system performance metrics in terms of scalability and timeliness based on the proposed SD-IIoT system architecture. Because the data generated by the IIoT FDs follow the CoAP and WebSocket standards in the UDP and TCP data flows, these two generic data flows will be evaluated.

*A. System Setup*

A typical WSN setting is considered, where the star WirelessHART network in the control loops in distributed sites. Three types of wireless sensors are deployed in the process control loops: Class 4 temperature sensors with the 1 s update interval, Class 2 pressure sensors with the 500 ms update interval, and Class 1 motor speed sensor with the 50 ms update interval. Each monitoring site has 6 temperature sensors, 6 pressure sensors, and 6 motor sensors. We assume the time slotted data transmission process is used by the WirelessHART FDs with the prioritized data transmission order by the QoS groups and sequence numbers of the FDs. The GW can therefore obtain the process values (PVs) sequentially from the surrounding FDs. All CoAP messages need to be confirmable (CON) in this case and the regular CoAP CON packet size is less than 100 bytes and the ACK packet size is 46 bytes. The entire network architecture follows the diagram in Fig. 2, where the Floodlight virtual SDN controller on the Mininet 2.1 virtual machine is used and one OVS is connected with one FD-GW of a field WSN. This host is communicating with a remote server controlled by the SDN controller. Corresponding to the three QoS groups, there are 3 queues created by *ovs-vsctl* tool at the Floodlight controller with the QoS module where each queue at OVS has a 100 Mbps bandwidth capacity. In addition, in order to compare the performance with the classical SC solution, we collect the results by using an Amazon AWS as an SC server without any QoS controller.

*B. Results*

We first evaluate the timeliness of CoAP-based protocol following a common data flow pattern of FDs, where the CoAP server is set in the observation mode, and a CoAP client on the GW communicates with the CoAP server on the FD for sensing data updates.

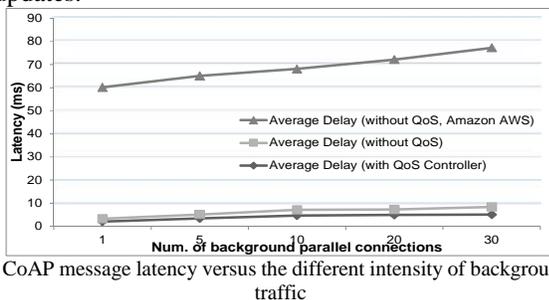
Fig. 3. CoAP message latency versus the different intensity of background UDP traffic

From Fig. 3, the classical SC solution on the public Amazon AWS has the worst performance. For the controlled industrial network solutions, we can see that SDN can significantly reduce the latency by 30%-38%, where the increasing number of parallel UDP traffic will increase the latency of the CoAP data flow but the average delay with QoS controllers only slightly increase when the parallel connection is over 10. This is due to the high priority and reliability for the CoAP packets set in the OVS queue by the QoS policies. Moreover, the background UDP traffic suffers the increasing data loss while the CoAP messages can still keep the 100% success rate, where this is because of the QoS mechanism in the CoAP messages as well as the policy set by the QoS controller. In this sense, if the network delay dominates the timeliness factors, the CoAP data flows can meet the QoS requirements of QoS groups mentioned in Section III.D.

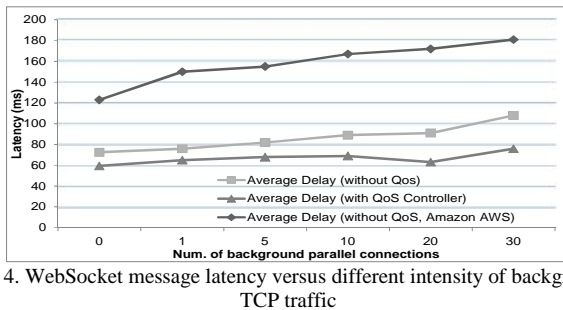
Fig. 4. WebSocket message latency versus different intensity of background TCP traffic

The latency performance of WebSocket data flows with and without SDN QoS controllers is shown in Fig. 4, where the flows with SDN QoS has steady latency around 60 ms when the number of background parallel TCP traffic is equal to or less than 20. Overall, with the SDN QoS controller, the latency in the controlled industrial network can be reduced from 17% to 44%. The Amazon AWS based solution has the largest delay, so it is not suitable for the safety-critical applications. In Fig. 4, the latency of data flows with the QoS controller becomes high when the number of parallel connections is 30. The reason of this phenomenon is that the background TCP traffic suffers from high connection timeouts and retransmission rates, which are brought by the TCP flow control mechanisms. This also explains why in the same condition WebSocket data flows have much higher latency than that of CoAP data flows which have much less control overhead. The results in Fig. 4 show the QoS policies can keep the latency in the desirable level even the background traffic is high.

In summary, from the aforementioned discussions, we can see that with the SDN QoS controller, the dynamic QoS control can be implemented on the SD-SC which provides a flexible and adaptive software definable framework to manage the important network performance in the IIoT. In this sense, the SD-IIoT performance together with other key components mentioned in Section III can be solved in a holistic way.

V. CONCLUSION

This paper intends to preliminarily address some key issues in a holistic IIoT system. An SD-IIoT architecture is proposed to address the essential requirements of generic IIoT applications, which can lead to an effective IIoT system design. Moreover, in order to achieve the timing requirements of safety-critical applications, an IIoT system needs to be implemented in a controlled and configurable environment. In the future work, we will analyze the SD-IIoT from the big data perspective employing additional deterministic networking techniques.